\documentclass[letterpaper]{article} 
\usepackage{aaai2026}  
\usepackage{times}  
\usepackage{helvet}  
\usepackage{courier}  
\usepackage[hyphens]{url}  
\usepackage{graphicx} 
\def\UrlFont{\rm}  
\usepackage{natbib}  
\usepackage{caption} 
\usepackage{algorithm}
\usepackage{algorithmic}
\usepackage{graphicx}
\usepackage{booktabs}
\usepackage{algorithm,algorithmic}
\usepackage{algorithm}
\usepackage{amsmath,amsfonts}
\usepackage{multirow}
\usepackage[table]{xcolor}
\usepackage{url}
\usepackage{newfloat}
\usepackage{listings}
\DeclareCaptionStyle{ruled}{labelfont=normalfont,labelsep=colon,strut=off} 
\floatstyle{ruled}
\newfloat{listing}{tb}{lst}{}
\floatname{listing}{Listing}
\title{IO-RAE: Information-Obfuscation Reversible Adversarial Example for Audio Privacy Protection}
\author{
    Jiajie Zhu\textsuperscript{\rm 1},
    Xia Du\textsuperscript{\rm 1}\thanks{Xia Du is the corresponding author.},
    Xiaoyuan Liu\textsuperscript{\rm 1},
    Jizhe Zhou\textsuperscript{\rm 2},
    Qizhen Xu\textsuperscript{\rm 1},
    Zheng Lin\textsuperscript{\rm 3},
    Chi-Man Pun\textsuperscript{\rm 4}
}
\affiliations{
    \textsuperscript{\rm 1}School of Computer and Information Engineering, Xiamen University of Technology, Xiamen, China\\
    \textsuperscript{\rm 2}School of Computer Science,
Engineering Research Center of
Machine Learning and Industry
Intelligence, Sichuan University
, Chengdu, China\\
    \textsuperscript{\rm 3}Department of Electrical and
Electronic Engineering, University of Hong Kong, Pok Fu Lam, Hong Kong
SAR, China\\
    \textsuperscript{\rm 4}Department of Computer and
Information Science, University of
Macau
Macau, China\\

    cosmics36@163.com, duxia@xmut.edu.cn, 2322071027@stu.xmut.edu.cn, jzzhou@scu.edu.cn, qzxu@xmut.edu.cn, linzheng@eee.hku.hk, cmpun@umac.mo
}

\usepackage{bibentry}

\begin{document}

\maketitle

\begin{abstract}
The rapid advancements in artificial intelligence have significantly accelerated the adoption of speech recognition technology, leading to its widespread integration across various applications. However, this surge in usage also highlights a critical issue: audio data is highly vulnerable to unauthorized exposure and analysis, posing significant privacy risks for businesses and individuals. This paper introduces an Information-Obfuscation Reversible Adversarial Example (IO-RAE) framework, the pioneering method designed to safeguard audio privacy using reversible adversarial examples. IO-RAE leverages large language models to generate misleading yet contextually coherent content, effectively preventing unauthorized eavesdropping by humans and Automatic Speech Recognition (ASR) systems. Additionally, we propose the Cumulative Signal Attack technique, which mitigates high-frequency noise and enhances attack efficacy by targeting low-frequency signals. Our approach ensures the protection of audio data without degrading its quality or usability. Experimental evaluations demonstrate the superiority of our method, achieving a targeted misguidance rate of 96.5\% and a remarkable 100\% untargeted misguidance rate in obfuscating target keywords across multiple ASR models, including a commercial black-box system from Google. Furthermore, the quality of the recovered audio, measured by the Perceptual Evaluation of Speech Quality score, reached 4.45, comparable to high-quality original recordings. Notably, the recovered audio processed by ASR systems exhibited an error rate of 0\%, indicating nearly lossless recovery. These results highlight the practical applicability and effectiveness of our IO-RAE framework in protecting sensitive audio privacy.
\end{abstract}

\section{Introduction}
Deep neural networks (DNNs) have revolutionized numerous domains due to their exceptional performance across various tasks \cite{DNNbrown2020language,fang2024automated,DNNdosovitskiy2020image,lin2024adaptsfl,duan2025llm,lin2025leo,yuan2025constructing,lin2025hierarchical,fang2024ic3m,lin2024fedsn,zhang2025lcfed,lin2024split,zhang2024satfed,yuan2024satsense,duan2025leed,lin2025hsplitlora}. Among these, Automatic Speech Recognition (ASR) has emerged as a critical application and extensively integrated into many aspects of daily life \cite{ASRhan2020contextnet,ASRsainath2021efficient,ASRng2021pushing}. However, the rapid and unregulated expansion of ASR technology has raised significant concerns regarding privacy and security. Malicious commercial entities increasingly exploit private voice recordings for profit-driven objectives, highlighting the urgent need to protect the integrity of audio data. 

\begin{figure}[t]
    \centering
    \includegraphics[width=1.0\linewidth]{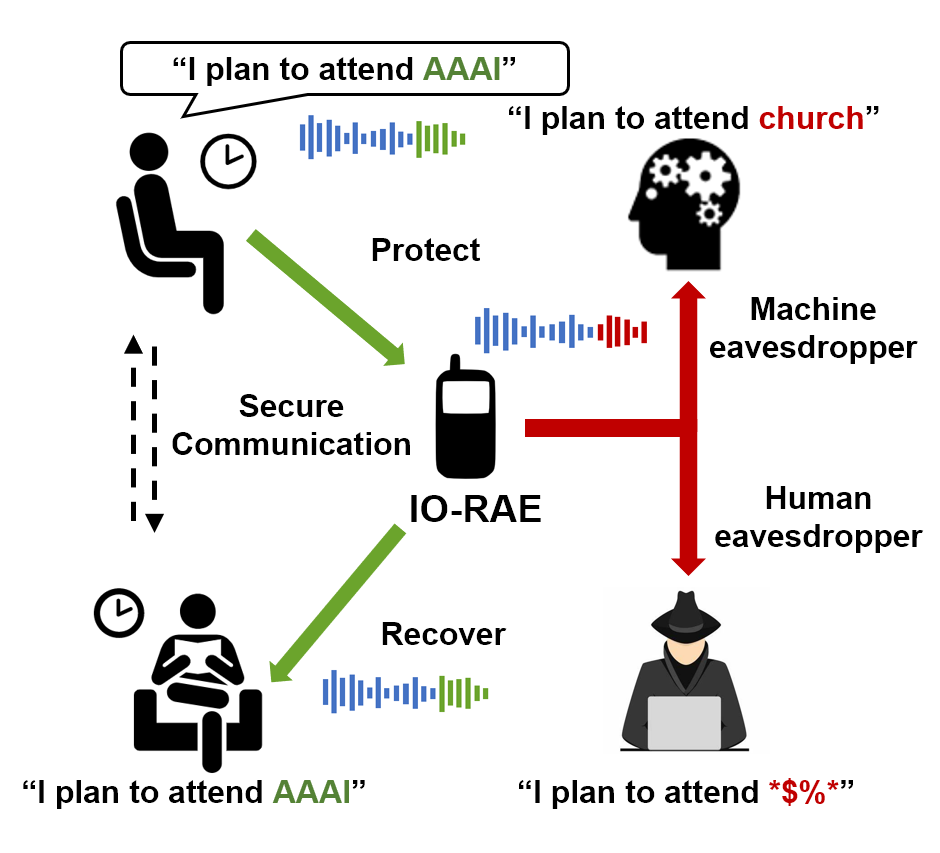}
    \caption{IO-RAE thwarts both human and machine eavesdroppers through the adversarial noise and can recover the original audio losslessly when authorized, ensuring secure communication.}
    \label{figure1}
\end{figure}

Recent research utlized the Data Anonymization techniques to protect user privacy by de-identifying or distorting personal features in speech signals, preventing ASR systems from extracting sensitive information. Chou \textit{et al.} \cite{chou2018multi} employed adversarial learning for multi-target voice conversion, which effectively removes speaker identity without the need for parallel data, facilitating anonymization. Champion \textit{et al.} \cite{champion2023anonymizing} developed a framework that isolates speaker-specific information by modifying acoustic features and embeddings. Feng \textit{et al.} \cite{feng2022enhancing} introduced domain-adaptive noise injection, which preserves privacy while mitigating the risk of identity disclosure. While these methods represent significant advancements in privacy protection, they still face challenges. Malicious attackers can still extract sensitive information using advanced DNNs, and the process of anonymization often leads to a degradation in speech clarity and intelligibility, limiting their practical applicability.



Given the limitations of existing anonymization methods, more effective approaches are needed to balance privacy protection and speech quality. In this paper, we apply Reversible Adversarial Examples (RAEs) \cite{liu2023unauthorized} to audio privacy protection: RAEs combine adversarial attacks with Reversible Data Hiding (RDH) so that crafted perturbations can mislead malicious DNNs and later be removed by authorized users to fully restore the original audio. However, current RAE designs face two major challenges: (1) enforcing imperceptible perturbations limits their obfuscation power, leaving sensitive information accessible to human listeners \cite{bigNoise,bigNoisebigNoise2,bigNoisebigNoise3}; (2) RDH capacity constraints require tight control of embedded data size, and creating adversarial patches to cover critical content is especially hard for ASR because locating key timestamps in audio is nontrivial \cite{RAE-VWP}.

To address these issues, we propose the Information Obfuscation Reversible Adversarial Example (IO-RAE) framework (Fig. \ref{figure1}). IO-RAE uses alignment techniques to locate sensitive timestamps and applies targeted voice attenuation together with adversarial patches to those segments, improving obfuscation while preserving overall intelligibility. We further employ Large Language Models (LLMs) \cite{GPT,qwen2.5} to generate deceptive labels for patches, making ASR transcriptions plausibly coherent, and introduce Cumulative Signal Attack (CSA), a perturbation optimization that reduces high-frequency harshness by accumulating signal energy into lower-frequency regions.

In summary, the main contributions of this paper are as follows:

$\bullet $ We proposed the IO-RAE for audio privacy protection, designed to prevent both human and machine eavesdroppers from extracting key information from the protected audio. To the best of our knowledge, this approach represents the first application of RAE in the audio domain.

$\bullet $ We proposed an LLM-based target generation method that replaces traditional predetermined targets and utilizes the greedy algorithm to identify the optimal target. This method preserves the coherence and plausibility of the sentences while substantially enhancing attack efficacy.

$\bullet $ We proposed the CSA method for noise preprocessing. CSA can smooth the harsh perturbations by suppressing high-frequency signals and maintain the efficacy of the attack by boosting low-frequency signals.

\section{Background}
\subsection{Adversarial Attack}
Adversarial attacks pose a fundamental threat to deep learning models, with their vulnerability validated across multimodal tasks. Szegedy \textit{et al.} \cite{szegedy2013intriguing} first revealed that imperceptible perturbations can significantly alter the predictions of image classification models. Goodfellow \textit{et al.} \cite{goodfellow2014explaining} then introduced the Fast Gradient Sign Method (FGSM), a simple yet effective one-step attack defined as:
\begin{equation}
\label{eq:eq0}
x_{adv}=x+\epsilon \cdot sign(\nabla _{x}\ell (x,y)),
\end{equation}
where $x$ is the input, $y$ the true label, and $\epsilon$ the perturbation scale. Since then, numerous attacks \cite{z1,z2,z3} have extended beyond image tasks.

In the audio domain, Carlini and Wagner \cite{CW} pioneered targeted attacks by minimizing the CTC loss \cite{CTC} while constraining perturbation magnitude, enabling adversarial speech to be transcribed as arbitrary phrases. Later, Yakura \textit{et al.} \cite{yakura2018robust} modeled environmental factors—such as room acoustics and noise—for robust over-the-air attacks. Qin \textit{et al.} \cite{Qin} introduced a psychoacoustic-aware framework that leverages auditory masking to create imperceptible yet effective adversarial audio, integrating human hearing models into the attack process.

\begin{figure*}[t]
    \centering
    \includegraphics[width=1\linewidth]{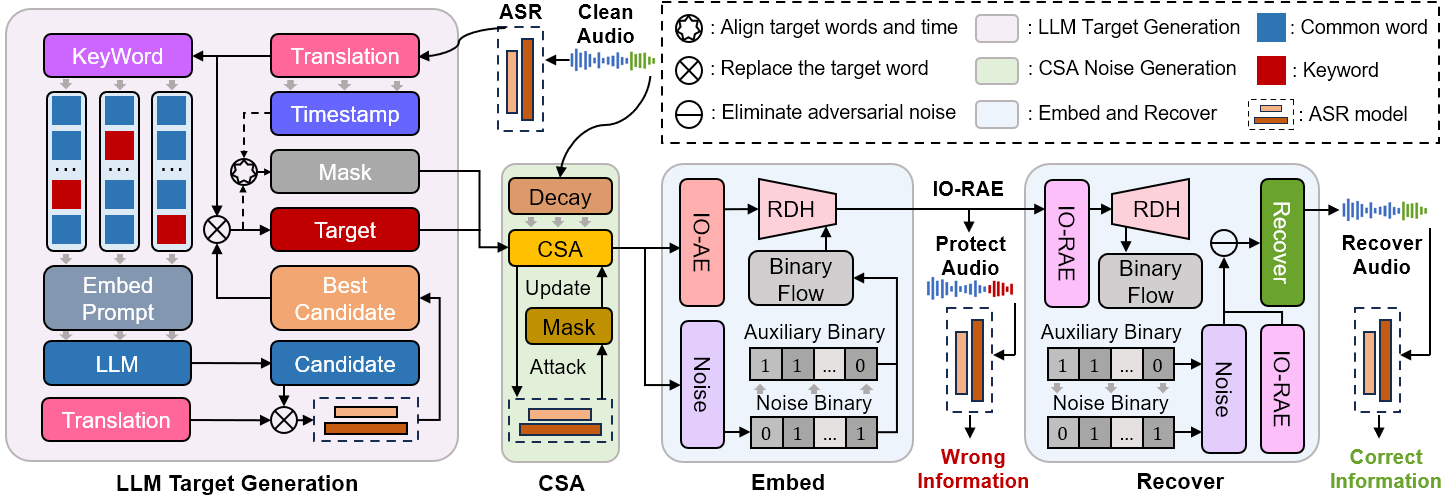}
    \caption{An overview of the IO-RAE framework.}
    \label{figure2}
\end{figure*}

\subsection{Reversible Adversarial attack}
Research on reversible adversarial attacks~\cite{zhang2025robust,zhang2025state} has primarily focused on the protection of image data~, while their application in the domain of audio data remains underexplored. Liu \textit{et al.} \cite{liu2023unauthorized}  first introduced the concept of reversible adversarial attacks by combining reversible data hiding techniques with adversarial examples, pioneering this novel attack methodology. Subsequently, Xiong \textit{et al.} \cite{xiong2023black} extended this work by applying reversible adversarial attacks to black-box scenarios, incorporating ensemble modeling techniques to demonstrate the potential of reversible adversarial examples across different models. Meanwhile, Zhang \textit{et al.} \cite{zhang2022self} proposed a method based on RGAN, which replaced reversible data hiding techniques. Specifically, they efficiently generated adversarial examples through an attack encoder network and reversed these examples effectively using a recovery decoder network, achieving efficient restoration of images. Despite its outstanding performance in recovering the original state of images, this method shows limited effectiveness when applied to adversarial tasks against black-box models. More recently, the DP-RAE framework \cite{dp-rae} improved attack success rates in black-box scenarios through a Dual-Phase design. In this paper, we propose the method that combines LSB \cite{LSB} with adversarial examples to achieve effective protection of audio data.

\section{Proposed Method}
\subsection{Overview}
In this section, we introduce the IO-RAE framework, which generates adversarial perturbations through the CSA to obfuscate critical information. As shown in Fig . \ref{figure2}, IO-RAE consists of three components: LLM Target Generation, adversarial perturbation creation through the CSA, and Reversible Perturbation Embedding and Recovery. Initially, IO-RAE utilizes KeyBERT \cite{keybert} to extract key information from the correct transcription. These keywords are then combined with prompts and fed into an LLM to generate logically consistent replacement words. Subsequently, the Montreal Forced Aligner MFA \cite{mfa} technique aligns the text and audio to obtain corresponding masks. Audio attenuation and perturbation generation are applied to these masked regions, with CSA to enhance low-frequency. Finally, auxiliary information and the perturbation matrix are embedded into the adversarial audio, forming the IO-RAE. During the recovery process, RDH extracts and decodes the perturbation matrix, effectively removing the adversarial perturbations.

\subsection{LLM Target Generation}
Unlike methods that generate target phrases based on fixed phrases or by maximizing word vector distance, we utilize LLMs to generate candidate replacement words. This approach avoids the confusion caused by words with the farthest vector distances and enhances diversity in the generated phrases. We denote the target transcription as $y = \left \{ y_{0}, y_{1}, ..., y_{n} \right \} $, where $y$ is initially set as the correct transcription and $y_{i}$ represents the $i_{th}$ word in the transcription. To achieve the goal of quoting out of context, it is necessary to identify a subset $y^{*} \subseteq y$ that can replace $y$ and effectively distort the original meaning. Therefore, we employ KeyBERT to identify the most influential words in $y$ to derive $y^{*}$ and utilize LLM to generate candidate words. This generates candidate words related to $y^{*}$. Moreover, the difficulty of misleading different classes in image recognition tasks varies, a phenomenon also observed in ASR systems. This variability is determined by the value of the loss function $\ell $ between the input and the target $y$:
\begin{equation}
\label{eq:eq1}
    \ell\left ( x,\delta ,y \right ) = \ell_{{\scriptsize CTC}} \left ( f(x+\delta),y \right),
\end{equation}
where $\delta$ represents the adversarial noise and $x$ denotes original audio. Therefore, we can conclude that there exists an optimal replacement set $y_{*}$ among the numerous candidate words that can mislead the ASR system at minimal cost. To efficiently determine $y^{*}$, we employ a greedy algorithm to identify the best replacement word for each position in $y^{*}$. Subsequently, we use the timestamps for each word provided by the MFA to generate the positional mask $\mathbb{M}$ for the perturbation $\delta$. The overall algorithm is detailed in Algorithm \ref{alg:odi}.

\begin{algorithm}[h]
\renewcommand{\algorithmicrequire}{\textbf{Input:}}
	\renewcommand{\algorithmicensure}{\textbf{Output:}}
    \caption{LLM Target Generation}
    \label{alg:odi}
    \begin{algorithmic}[1]
    
        \REQUIRE Original audio $x$.\\
        \ENSURE{Output:} Target transcription $y$, location mask $\mathbb{M}$.\\
        \renewcommand{\algorithmicrequire}{\textbf{Initialize:}}
       \REQUIRE $y = f\left ( x \right ) $, $y^{*}=\left \{  \right \} $, $\mathcal{L} = \left \{  \right \}$, $\mathbb{M}\gets  0\left [  shape(x)\right ]$\\

            \STATE Obtain the keywords: $y^{*}={\tiny \mathrm{KeyBERT } }(y)$;
            \STATE Utilize MFA to align the $x$ and $y$;
		\FOR{$i = 0 \rightarrow len(y^{*})-1$}
		    \STATE Embedding $y_{i}^{*}$ into prompts;
                \STATE Obtain the candidate lexicon $cl$: $cl$ = LLM(prompts);
                \FOR{$j = 0 \rightarrow len(cl)-1$}
                    \STATE Replace $y_{\;idx(y_{i}^{*} \; in \; y)}$ with $cl_{j}$;
                    \STATE Calculate the loss $\ell $ with Eq. \ref{eq:eq1};
                    \STATE Append $\ell $ to $\mathcal{L}$;
                \ENDFOR
                \STATE Find the index $min$ with the smallest value in $\mathcal{L}$;
                \STATE Replace $y_{\;idx(y_{i}^{*} \; in \; y)}$ with $cl_{min}$;
                \STATE Get the start and end time of the $y_{\;idx(y_{i}^{*} \; in \; y)}$;
                \STATE $\mathbb{M}\left [ start:end \right ] =1$;
		\ENDFOR
        \STATE \textbf{return} $y$, $\mathbb{M}$
	\end{algorithmic} 

\end{algorithm} 

The optimal target transcription $y$ and the location mask $\mathbb{M}$ found using the LLM and greedy-based concepts will be utilized in the subsequent CSA attack to generate targeted perturbations.

\subsection{Cumulative Signal Attack}
Audio data storage is unique as even short audio segments require extensive tensors for representation due to their high sampling rates, resulting in substantial data volumes in the full perturbation matrix. Given the constrained storage capacity of RDH technology, these perturbations must be compressed prior to embedding. Consequently, in CSA, we utilize composite sampling to reduce data size while preserving attack effectiveness. Moreover, CSA combats overfitting in audio attacks by preventing the generation of additional perturbations for words already successfully attacked and using cumulative signals to soften harsh noises.

The CSA method first decays the original audio content to ensure the imperceptibility of the protected material, then applies a composite sample operation on the matrix to ensure that the compressed gradient matrix retains its adversarial properties. Assuming the size of the composite sample is $s$ and the compressed matrix is $\mathcal{L} _{c}$. The value of the $i_{th}$ element in $\mathcal{L} _{c}$ can be expressed as:
\begin{equation}
\label{eq:eq2}
    \mathcal{L} _{c}[i]=\sum_{j=i*s}^{i*s+s} \nabla _{x}\ell ( x,\delta ,y)_{j}/s,
\end{equation}
then the CSA method replaces the original gradients in the interval $\left [ i:i+s \right ] $ with the averaged compressed gradient $\mathcal{L} _{c}[i]$. This adjustment allows the perturbation to align with the compressed gradients.

Subsequently, the CSA method optimizes the generated perturbations by enhancing the low-frequency components. After undergoing the Discrete Fourier Transform \cite{dftcooley1965algorithm}, the perturbation signal $\delta$ is represented in the frequency domain as:
\begin{equation}
\label{eq:eq3}
      \triangle (k)={\textstyle \sum_{n=0}^{N-1}} \delta \left [ n \right ] e^{-j\frac{2\pi kn}{N} },
\end{equation}
and the frequency domain of the cumulative signal $c$ corresponding to $\delta$ is:
\begin{equation}
\label{eq:eq4}
      C(k)= {\textstyle \sum_{m=0}^{N-1}} \delta [m]\left (  {\textstyle \sum_{n=m}^{N-1}} e^{-j\frac{2\pi kn}{N}}  \right ),
\end{equation}
for each fixed $m$, one obtains:
\begin{equation}
\label{eq:eq5}
      {\textstyle \sum_{n=m}^{N-1}} e^{-j\frac{2\pi kn}{N}}=e^{-mj\frac{2\pi kn}{N}} {\textstyle \sum_{n=0}^{N-1-m}} e^{-j\frac{2\pi kn}{N}}.
\end{equation}
When N is sufficiently large, Eq. \ref{eq:eq5} can be approximated as:
\begin{equation}
\label{eq:eq6}
      {\textstyle \sum_{n=0}^{N-1-m}} e^{-j\frac{2\pi kn}{N}} \approx \frac{1}{1-e^{-j\frac{2\pi k}{N} }}, 
\end{equation}
the relationship between $C(k)$ and $\triangle (k)$ can be expressed as:
\begin{equation}
\label{eq:eq7}
      C(k) \approx \triangle (k) \cdot \frac{1}{1-e^{-j\frac{2\pi k}{N} }}.
\end{equation}
Thus, the cumulative operation on the perturbed signal enhances the low-frequency components while suppressing the high-frequency components.  The detailed derivation is provided in the Supplementary Material.

To prevent overfitting to specific words during the perturbation generation process, the CSA method evaluates the attack success rate for each target replacement word before each iteration, updating the success matrix $\mathbb{S}$. If a word has already been successfully attacked, no additional perturbations are applied in subsequent iterations. The complete algorithm is presented in Algorithm \ref{alg:alg2}.

\begin{algorithm}[h]
\renewcommand{\algorithmicrequire}{\textbf{Input:}}
	\renewcommand{\algorithmicensure}{\textbf{Output:}}
\caption{Cumulative Signal Attack}
\label{alg:alg2}

\begin{algorithmic}[1]
    \REQUIRE Original audio $x$, target $y$, decay rate $\alpha$, mask $\mathbb{M}$,Iteration $I$, unit noise $\epsilon$, multiplicative factor $t$
    \ENSURE Adversarial audio $x_{adv}$, noise matrix $\delta$
    \renewcommand{\algorithmicrequire}{\textbf{Initialize:}}
    \REQUIRE $g_{0}=0$, $x_{adv}^{0}=\frac{1}{\alpha } \cdot x \odot \mathbb{M}$, CSA noise $\eta=0$, $\delta \gets 0[\text{shape}(x)]$, $\eta \gets 0[\text{shape}(x)]$, $\mathbb{S} \gets \mathbb{M}$, $\epsilon = \epsilon\cdot \frac{1+\alpha/5}{100 } $

    \FOR{$i = 0 \rightarrow I-1$}
        \STATE $y_{adv} = f(x_{adv}^{i} + \eta)$
        \STATE Verify if the attack succeeds on $y_{adv}$
        \STATE Update $\mathbb{S}$ based on the result of $y_{adv}$
        \STATE Calculate the loss for Eq.~\ref{eq:eq1} and obtain $g_{t+1}$
        \STATE $g_{t+1} = \mu \cdot g_{t} + \frac{g_{t+1}}{\| g_{t+1} \|_2}$
        \STATE Compress and update $g_{t+1}$ using Eq.~\ref{eq:eq2}
        \STATE $\delta = \delta + \epsilon \cdot \text{sign}(\mathbb{M} \odot \mathbb{S} \cdot g_{t+1})$
        \STATE $\delta = \text{Clip}_{\delta}^{t \cdot \delta}(\delta)$
        \STATE $\eta = {\frac{1}{\alpha } \cdot \textstyle \sum_{m=0}^{N-1}} \delta[m]$ \hfill $\triangleright$ Cumulative Signal
        \STATE $x_{adv}^{i+1} = x_{adv}^{i} + \eta$
    \ENDFOR
    \STATE \textbf{return} $x_{adv}, \delta$
\end{algorithmic}

\end{algorithm}

Subsequently, the adversarial examples and their corresponding perturbation matrices are employed in the next phase to construct reversible adversarial audio attacks. Notably, to facilitate information encoding and reduce storage overhead, the returned perturbation matrices correspond to the original matrices $\delta$ prior to the cumulative signal operation.
\subsection{Embed and Recover}
In the IO-RAE framework, we implement RDH techniques to embed additional data into adversarial examples. Initially, the perturbation matrix generated by the CSA method is encoded. Specifically, the perturbations are sampled at intervals of $s$, capturing the coefficients of the standard perturbation $\epsilon$, the message of $i_{th}$ element can be expressed as:
\begin{equation}
\label{eq:eq8}
   Mes_{i}=\frac{\sum_{j=i*s}^{i*s+s}\delta _{j}}{\epsilon \times s}.
\end{equation}

All valid perturbation matrices are sequentially traversed, and their correspondi ng $Mes_{i}$ are aggregated to form $Mes$, resulting in the perturbation data stream. This stream is then combined with auxiliary information—including the starting position of the perturbation, the standard perturbation magnitude, and the decay rate. The final composite information is embedded into the adversarial audio to ensure its adversarial properties are preserved.

During the audio recovery phase, the embedded information is first extracted using RDH technique. With the extracted $Mes$ and auxiliary data, the perturbation matrix and its corresponding embedding positions are reconstructed in reverse.Subsequently, the forward process of CSA is reapplied to the perturbed matrix to obtain the CSA noise $\eta$. Finally, the original audio is effectively restored by progressively removing the perturbations from the adversarial audio.
\section{Experiments}
\subsection{Experiment setup}
\textbf{Datasets.} We randomly selected 200 accurately transcribed audio samples from each of Mozilla Common Voice \cite{Common}, TIMIT \cite{TIMIT}, and LibriSpeech \cite{librispeech} to validate the method’s effectiveness across diverse data sources.

\noindent %
\textbf{Environment.} 
We used the widely adopted DeepSpeechV3 \cite{deepspeechV3} as the primary target model due to its relevance in speech recognition. For transferability evaluation, we tested on Conformer \cite{conformer}, Whisper \cite{Whisper}, Wenet \cite{wenet}, SenseVoice \cite{funasr}, and Wav2vec \cite{wav2vec}. All experiments were implemented in PyTorch and conducted on an NVIDIA A40 GPU.

\noindent %
\textbf{Evaluation Metrics.}
To assess audio quality and intelligibility, we used SNR, SISDR, STOI, and PESQ. For attack effectiveness, we evaluated targeted success rate (TSR), targeted word error rate (TWER), untargeted success rate (USR), and untargeted word error rate (UWER).

\noindent %
\textbf{Parameter Setting.}
We set unit noise $\epsilon$ to $1/5$ and the multiplicative factor $t$ to 15, implying each perturbation requires only 5 bits. The composite sample size $s$ was set to 2, balancing storage and effectiveness. Audio attenuation $\alpha$ was set at 5 per 0.1 of maximum signal, capped at 30. The number of attack iterations $I$ was fixed at 200. And for target word generation, we employed Qwen2.5-VL-7B.

\subsection{Attack and Recover ability}
In this section, we demonstrate the performance of IO-RAE in both attack and recovery on three major datasets: Common, TIMIT, and LibriSpeech, specifically targeting the DeepSpeechV3 model. The evaluation dimensions primarily include audio quality and attack performance.

\begin{figure}[h]
    \centering
    \includegraphics[width=0.9\linewidth]{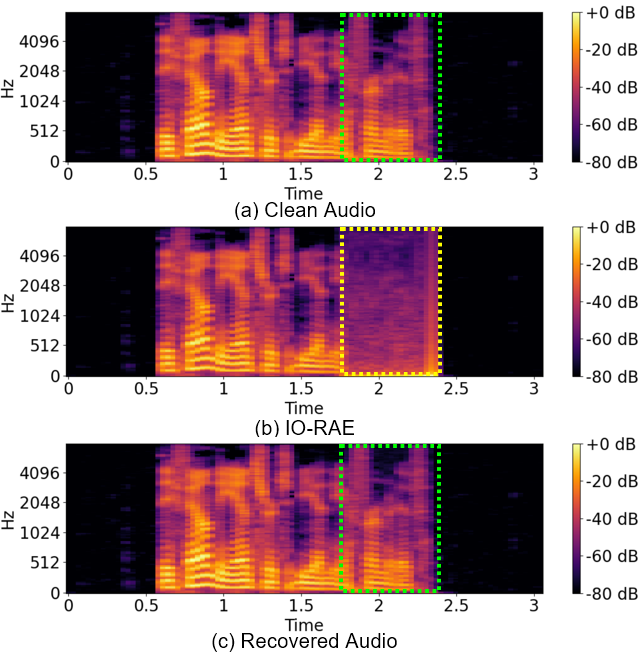}
    \caption{(a), (b), and (c) represent the Mel spectrogram of clean audio, IO-RAE, and recovered audio, respectively.}
    \label{figure3}
\end{figure}

Regarding attack performance, as shown in Table \ref{tab:table1}, IO-RAE exhibits high target misleading capability across multiple datasets. Notably, on the LibriSpeech dataset, the TSR reached 96.5\%, while the TWER was only 0.21\%, indicating minimal deviation between the transcription results and the target output even when the target attack was not fully successful. Therefore, the ASR system struggled to extract accurate information from the adversarial audio. As depicted in the Mel spectrogram \cite{mel} in Fig . \ref{figure3}, the attacked audio regions were fully protected through CSA.

In terms of recovery, Fig . \ref{figure3} visually illustrates that the recovered audio was free from misleading perturbations and perfectly restored to the clean sample state. Experimental results in Table \ref{tab:table1} further confirm that the recovered audio exhibited exceptional quality in various tests, with a PESQ score of 4.45, close to the theoretical maximum of 4.5, objectively validating the superiority of the restored audio. Other critical audio quality metrics, such as SNR, SISDR, and STOI, also showed significant improvements. Re-transcription of the recovered audio using DeepSpeechV3 achieved a TSR of 0\%, effectively restoring the original information and proving the effectiveness and practicality of our method.

\begin{table}[h]
\centering
\tiny
\renewcommand{\arraystretch}{1}
\resizebox{1\linewidth}{!}{
\centering
\begin{tabular}{lcccc|cc}
\toprule[0.15em]
\multirow{2}{*}{\begin{tabular}[c]{@{}c@{}}Source:\\ Common\end{tabular}} & \multicolumn{4}{c|}{Audio Quality} & \multicolumn{2}{c}{Attack Performance} \\
& SNR & SISDR & STOI & PESQ & TSR & TWER  \\ \toprule[0.1em] 

IO-AE   & 1.8   & -0.22 & 0.83 & 2.6  & 96.0 & 0.6   \\
IO-RAE  & 1.8   & -0.22 & 0.83 & 2.6  & 96.5 & 0.55  \\
Recover & \textcolor{red}{53.06} & \textcolor{red}{53.06} & \textcolor{red}{0.99} & \textcolor{red}{4.38} & \textcolor{red}{0}    & \textcolor{red}{16.18} \\

\midrule

\multirow{2}{*}{\begin{tabular}[c]{@{}c@{}}Source:\\ TIMIT\end{tabular}} & \multicolumn{4}{c|}{Audio Quality} & \multicolumn{2}{c}{Attack Performance} \\
& SNR & SISDR & STOI & PESQ & TSR & TWER  \\ \toprule[0.1em] 

IO-AE   & -12.37 & -15.35 & 0.78 & 2.41 & 96.0 & 1.08  \\
IO-RAE  & -12.37 & -15.35 & 0.78 & 2.41 & 95.0 & 1.36  \\
Recover & \textcolor{red}{42.76}  & \textcolor{red}{42.76}  & \textcolor{red}{0.99} & \textcolor{red}{4.28} & \textcolor{red}{0}    & \textcolor{red}{17.31} \\

\midrule

\multirow{2}{*}{\begin{tabular}[c]{@{}c@{}}Source:\\ Librispeech\end{tabular}} & \multicolumn{4}{c|}{Audio Quality} & \multicolumn{2}{c}{Attack Performance} \\
& SNR & SISDR & STOI & PESQ & TSR & TWER  \\ \toprule[0.1em] 

IO-AE   & 3.37  & 1.99  & 0.89 & 3.02 & 96.5 & 0.21 \\
IO-RAE  & 3.28  & 1.91  & 0.89 & 3.02 & 96.5 & 0.21 \\
Recover & \textcolor{red}{54.75} & \textcolor{red}{54.75} & \textcolor{red}{0.99} & \textcolor{red}{4.45} & \textcolor{red}{0}    & \textcolor{red}{10.2} \\

\bottomrule[0.15em]
\end{tabular}}

\caption{SNR (dB), SISDR (dB), STOI, PESQ, TSR (\%) and TWER (\%) of the adversarial audio (IO-AE and IO-RAE) and recover audio. Recovered results are highlighted in \textcolor{red}{red}.}
\label{tab:table1}
\end{table}

\subsection{Robustness evaluation}
In real-world applications, input audio typically undergoes a series of defense methods aimed at mitigating the impact of adversarial noise on the model. Therefore, ensuring the robustness of adversarial samples is crucial. To evaluate the robustness of IO-RAE, we applied various defense methods, including Average Smoothing (AS) \cite{AS}, Audio Turbulence (AT) \cite{AT}, Low Pass Filter (LPF) \cite{LPF}, Median Smoothing (MS) \cite{MS&QT}, Quantization (QT) \cite{MS&QT}, MP3-V and OPUS \cite{OPUS} to the adversarial samples.

As shown in Table \ref{tab:table2}, despite the varying degrees of impact these defense methods had on targeted attacks, the adversarial examples produced by IO-RAE consistently achieve near-perfect mislead rates close to 100\% across all datasets, effectively obstructing the accurate interpretation of the original information. This robustness can be attributed to IO-RAE's reversible adversarial attack nature, which allows for stronger perturbations without compromising overall audio quality. Consequently, IO-RAE demonstrates strong robustness and reliability in real-world applications.

\begin{table}[t]
\tiny
\renewcommand{\arraystretch}{1}

\resizebox{\linewidth}{!}{
\centering
\begin{tabular}{lccccccc}
\toprule[0.15em]
\multirow{2}{*}{\begin{tabular}[c]{@{}c@{}}Source:\\ Common\end{tabular}} & \multicolumn{7}{c}{Defense method} \\
\cmidrule(l){2-8}
& AS & AT & LPF & MS & QT & MP3-V & OPUS \\ \toprule[0.1em] 

TSR $\uparrow$  & 8.0   & 45.0  & 26.0  & 13.5  & \textbf{59.0}  & 25.0  & 19.5  \\
TWER $\downarrow$ & 22.7  & 9.45  & 13.79 & 24.17 & \textbf{6.89}  & 13.39 & 15.99 \\
USR $\uparrow$  & \textbf{100}   & \textbf{100}   & \textbf{100}   & \textbf{100}   & 98.5  & \textbf{100}   & \textbf{100}   \\
UWER $\uparrow$ & 24.13 & 18.33 & 19.29 & \textbf{29.19} & 17.14 & 18.31 & 19.86 \\

\midrule[0.1em]

\multirow{2}{*}{\begin{tabular}[c]{@{}c@{}}Source:\\ TIMIT\end{tabular}} & \multicolumn{7}{c}{Defense method} \\
\cmidrule(l){2-8}
& AS & AT & LPF & MS & QT & MP3-V & OPUS  \\ \toprule[0.1em] 

TSR $\uparrow$  & 4.5   & 32.0  & 38.0  & 20.5  & 34.5  & 19.5  & 9.0   \\
TWER $\downarrow$ & 27.27 & 18.09 & 11.72 & 23.82 & 13.17 & 16.63 & 19.28 \\
USR $\uparrow$  & 99.5  & 99.5  & \textbf{100}   & \textbf{100}   & \textbf{100}   & 99.0  & 99.5  \\
UWER $\uparrow$ & 28.98 & 25.75 & 20.45 & \textbf{30.31} & 23.31 & 21.66 & 22.27 \\

\midrule[0.1em]

\multirow{2}{*}{\begin{tabular}[c]{@{}c@{}}Source:\\ Librispeech\end{tabular}} & \multicolumn{7}{c}{Defense method} \\
\cmidrule(l){2-8}
& AS & AT & LPF & MS & QT & MP3-V & OPUS \\ \toprule[0.1em] 

TSR $\uparrow$  & 10.5  & 57.5  & 50.5  & 12.5  & \textbf{68.5}  & 32.0  & 30.0  \\
TWER $\downarrow$ & 12.69 & 4.58  & 5.11  & 13.9  & \textbf{3.1}   & 8.13  & 8.07  \\
USR $\uparrow$  & \textbf{99.5}  & \textbf{99.5}  & \textbf{99.5}  & \textbf{99.5}  & 99.0  & \textbf{99.5}  & \textbf{99.5}  \\
UWER $\uparrow$ & 14.01 & 10.69 & 11.03 & \textbf{16.91} & 10.43 & 11.77 & 11.38 \\

\bottomrule[0.15em]
\end{tabular}}
\caption{TSR (\%), TWER (\%), USR (\%) and UWER (\%) of IO-RAE under various defense methods.}
\label{tab:table2}
\end{table}

\subsection{Transferability evaluation}
Since the specific architecture used by an intruder's model is often unknown, it is necessary to verify whether IO-RAE can protect critical target information across different models. In the field of image attacks, transferable perturbations can be generated by computing ensemble gradients due to the similarity in model architectures, effectively attacking unknown models. However, in the audio attack domain, significant differences in model architectures make cross-model transfer much more difficult and challenging. IO-RAE utilizes CSA to effectively conceal critical information, achieving cross-model protection. We selected several models with vastly different architectures to evaluate the transferability of IO-RAE, including Wenet, Conformer, SenseVoice, Whisper, and Wav2vec.

As shown in Table \ref{tab:table3}, IO-RAE was able to effectively protect the key content across various models, preventing multiple models from retrieving the correct original information. This demonstrates that IO-RAE possesses strong cross-model information security capabilities, maintaining the protection of critical information even when faced with different architectures.
\begin{table}[H]
\tiny
\renewcommand{\arraystretch}{1}

\resizebox{\linewidth}{!}{
\centering
\begin{tabular}{lcc|cc|cc}
\toprule[0.15em]
Source: & \multicolumn{2}{c|}{Common}& \multicolumn{2}{c|}{TIMIT}& \multicolumn{2}{c}{ibrispeech} \\
model & USR $\uparrow$  & UWER $\uparrow$  & USR $\uparrow$  & UWER $\uparrow$  & USR $\uparrow$  & UWER $\uparrow$  \\ \toprule[0.1em] 
Wenet      & 91.5 & 18.25 & 89.5 & 17.48 & 85.5 & 10.63 \\
Conformer  & 94.0 & 19.72 & 98.0 & 21.61 & 92.0 & 11.38 \\
SenseVoice & 96.0 & 21.06 & 98.0 & 19.85 & 93.5 & 13.23 \\
Whisper    & 99.5 & \textbf{29.01} & \textbf{100}  & \textbf{29.59} & \textbf{99.5} & \textbf{29.26} \\
Wav2vec    & \textbf{100}  & 22.67 & \textbf{100}  & 22.62 & 99.0 & 12.22 \\

\midrule[0.1em]
\end{tabular}}
\caption{USR (\%) and UWER (\%) of IO-RAE targeting various ASRs, the IO-RAE is generated by DeepspeechV3.}
\label{tab:table3}
\end{table}

\subsection{Commercial model attack}
To further validate the effectiveness of our model in real-world scenarios, we conducted attack experiments targeting the Google API \footnote[1]{\url{https://cloud.google.com/speech-to-text}}, the commercial ASR system. Our objective was to deceive the Google API with IO-RAE, ensuring that it failed to accurately extract the correct information from audio signals. We prepared 50 voice samples from the Common Voice dataset, all of which were initially translated accurately by the API. Subsequently, we generated the corresponding IO-RAE to deceive the API. Thanks to the disturbances introduced by CSA, the IO-RAE completely safeguarded the key content, achieving a 100\% attack success rate. In addition, we also recovered the audio from IO-RAE, and the recovered audio could be correctly recognized by Google API again, which means that the recovered audio wipes out the adversarial perturbation and restores the audio quality. We uploaded the corresponding audio data to an anonymous website \footnote[2]{\url{https://sites.google.com/view/io-rae/io-rae/}}, allowing for playback and download to audition the attack and restoration effects.

\begin{figure}[h]
    \centering
    \includegraphics[width=1.0\linewidth]{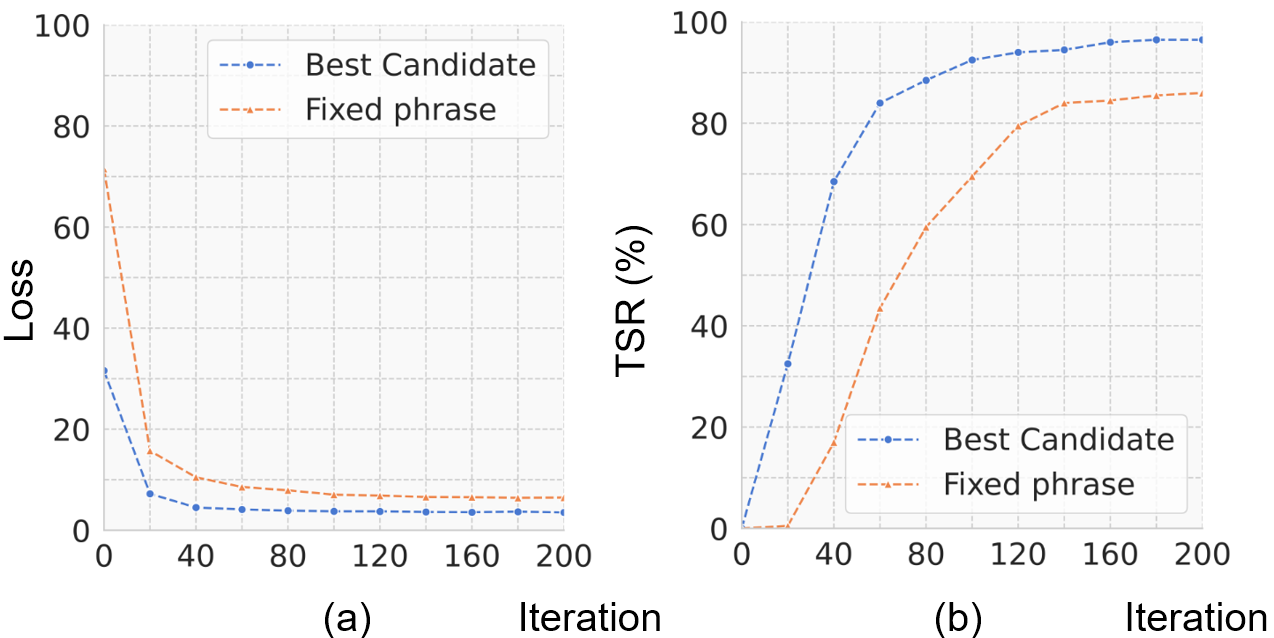}
    \caption{(a) represents the CTC loss and (b) represents the TSR (\%) of two strategies in the Common dataset.}
    \label{figure4}
\end{figure}

\subsection{Ablation study}
\begin{table*}[t]
\renewcommand{\arraystretch}{1}

\footnotesize
\resizebox{\linewidth}{!}{%
\begin{tabular}{lcccc|cccc|ccc|ccc}
\toprule[0.15em]
\textbf{Source:}& \multicolumn{8}{c}{Composite Sample}& \multicolumn{6}{c}{Unit Noise}\\
\cmidrule(l){2-15}
\textbf{Common}& \multicolumn{4}{c|}{IO-AE}& \multicolumn{4}{c|}{IO-RAE}& \multicolumn{3}{c|}{IO-AE} & \multicolumn{3}{c}{IO-RAE}\\
  & $\;s=2\;$ & $\;s=3\;$ & $\;s=4\;$ & $\;s=5\;$ & $\;s=2\;$ & $\;s=3\;$ & $\;s=4\;$ & $\;s=5\;$ & $\epsilon=1/4$ & $\epsilon=1/5$ & $\epsilon=1/6$ & $\epsilon=1/4$ & $\epsilon=1/5$ & $\epsilon=1/6$ \\ \midrule
  
TSR $\uparrow$ & 96.0    & 92.0 & 91.0  & 84.5 & 96.5  & 93.0  & 89.5  & 84.5 & 96.0  & 96.0  & 93.0  & 95.5  & 96.5  & 92.5  \\
TWER $\downarrow$ & 0.60  & 1.42 & 1.54  & 3.05 & 0.55  & 1.67  & 1.96  & 3.16 & 0.62  & 0.60  & 0.60  & 0.75  & 0.55  & 1.33  \\
USR $\uparrow$ & 100   & 100  & 100   & 100  & 100   & 100   & 100   & 100  & 100   & 100   & 100   & 100   & 100   & 100   \\
UWER $\uparrow$ & 16.21 & 16.2 & 16.27 & 16.8 & 16.16 & 16.33 & 16.27 & 16.8 & 16.08 & 16.21 & 16.21 & 16.08 & 16.16 & 16.22 \\

\midrule[0.1em]
\textbf{Source:}& \multicolumn{8}{c}{Composite Sample}& \multicolumn{6}{c}{Unit Noise}\\
\cmidrule(l){2-15}
\textbf{TIMIT}& \multicolumn{4}{c|}{IO-AE}& \multicolumn{4}{c|}{IO-RAE}& \multicolumn{3}{c|}{IO-AE} & \multicolumn{3}{c}{IO-RAE}\\
  & $\;s=2\;$ & $\;s=3\;$ & $\;s=4\;$ & $\;s=5\;$ & $\;s=2\;$ & $\;s=3\;$ & $\;s=4\;$ & $\;s=5\;$ & $\epsilon=1/4$ & $\epsilon=1/5$ & $\epsilon=1/6$ & $\epsilon=1/4$ & $\epsilon=1/5$ & $\epsilon=1/6$ \\ \midrule

TSR $\uparrow$ & 96.0  & 91.0 & 84.5  & 83.5  & 95.0  & 88.5 & 83.5 & 82.0  & 88.5  & 96.0  & 93.5 & 85.5  & 95.0  & 90.5  \\
TWER $\downarrow$ & 1.08  & 2.03 & 3.30  & 3.38  & 1.36  & 2.61 & 3.52 & 3.62  & 2.09  & 1.08  & 1.55 & 2.57  & 1.36  & 2.02  \\
USR $\uparrow$ & 100   & 100  & 100   & 100   & 100   & 100  & 100  & 100   & 100   & 100   & 100  & 100   & 100   & 100   \\
UWER $\uparrow$ & 17.74 & 17.6 & 18.07 & 17.92 & 17.74 & 17.90 & 18.1 & 18.08 & 17.59 & 17.74 & 17.40 & 17.89 & 17.74 & 17.52 \\

\midrule[0.1em]
\textbf{Source:}& \multicolumn{8}{c}{Composite Sample}& \multicolumn{6}{c}{Unit Noise}\\
\cmidrule(l){2-15}
\textbf{Librispeech}& \multicolumn{4}{c|}{IO-AE}& \multicolumn{4}{c|}{IO-RAE}& \multicolumn{3}{c|}{IO-AE} & \multicolumn{3}{c}{IO-RAE}\\
   & $\;s=2\;$ & $\;s=3\;$ & $\;s=4\;$ & $\;s=5\;$ & $\;s=2\;$ & $\;s=3\;$ & $\;s=4\;$ & $\;s=5\;$ & $\epsilon=1/4$ & $\epsilon=1/5$ & $\epsilon=1/6$ & $\epsilon=1/4$ & $\epsilon=1/5$ & $\epsilon=1/6$ \\ \midrule
  
TSR $\uparrow$ & 96.5  & 94.5  & 90.5  & 87.5 & 96.5  & 94.0  & 90.5  & 87.0 & 94.5  & 96.5  & 96.0  & 94.0  & 96.5  & 95.5  \\
TWER $\downarrow$ & 0.21  & 0.47  & 0.95  & 1.75 & 0.22  & 0.52  & 0.95  & 1.84 & 0.57  & 0.21  & 0.25  & 0.71  & 0.22  & 0.28  \\
USR $\uparrow$ & 100   & 100   & 100   & 100  & 100   & 100   & 100   & 100  & 100   & 100   & 100   & 100   & 100   & 100   \\
UWER $\uparrow$ & 10.22 & 10.34 & 10.44 & 10.70 & 10.22 & 10.34 & 10.44 & 10.70 & 10.34 & 10.22 & 10.23 & 10.39 & 10.22 & 10.26 \\

\bottomrule
\end{tabular}}
\caption{TSR (\%), TWER (\%), USR (\%), UWER (\%) of IO-RAE with different settings. The composite sample demonstrates the attack performance of IO-RAE with different composite sample sizes, and the unit noise represents the impact of adversarial noise of varying degrees on performance.}
\label{tab:table4}
\end{table*}

We initially explored the impact of the 'best candidate strategy' on attack performance. As illustrated in Fig . \ref{figure4}, a remarkable improvement in attack performance is observed when the 'best candidate strategy' replaces the 'fixed strategy'. This improvement occurs because a higher loss between target words and the input audio indicates a lower perceived match by the ASR, necessitating larger or more complex perturbations to coerce the ASR into outputting the target words. The 'best candidate strategy' effectively reduces this loss, thereby decreasing the required perturbation intensity and enhancing the TSR.

\begin{figure}[h]
    \centering
    \includegraphics[width=0.9\linewidth]{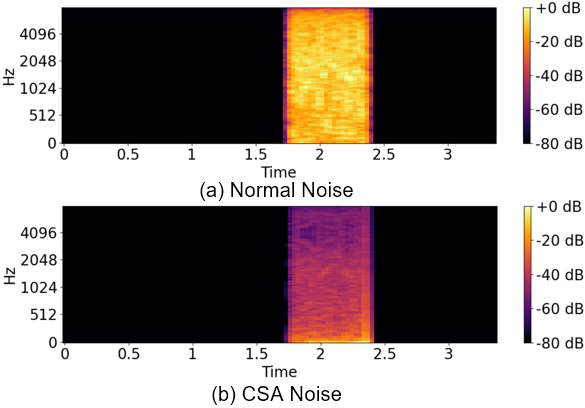}
    \caption{(a) represents the normal noise and (b) represents the CSA noise (ours).}
    \label{figure5}
\end{figure}

To evaluate CSA's effectiveness on sharp noise, we compared normal and CSA-generated noise. As shown in Fig. \ref{figure5}, spectral analysis indicates that normal noise retains high energy in mid-to-high frequencies, whereas CSA significantly suppresses energy in these bands, reducing harshness and producing a smoother noise profile.

We then examined two key factors: composite sample size and perturbation magnitude. As shown in Table \ref{tab:table4}, larger composite sizes reduce attack performance due to gradient averaging, which weakens the precision of perturbation direction. Conversely, overly small samples increase storage demands due to high information density (Table \ref{tab:table5}). Thus, a composite size of 2 achieves a balance between efficiency and performance. For perturbation magnitude, increasing scale consistently improves attack success. We found that a scale of $1/5$ offers the best trade-off between perturbation subtlety and attack effectiveness.

\begin{table}[H]
\tiny
\renewcommand{\arraystretch}{1}

\resizebox{\linewidth}{!}{
\centering
\begin{tabular}{lccccc}
\toprule[0.15em]
  & \multicolumn{5}{c}{Composite Sample} \\
\cmidrule(l){2-6}
Source & $s=1$ & $s=2$ & $s=3$ & $s=4$  & $s=5$\\ \midrule[0.1em]
Common      & 51,603 & 25,811 & 17,214 & 12,915 & 10,336 \\
TIMIT       & 47,627 & 23,823 & 15,889 & 11,921 & 9,541  \\
Librispeech & 42,296 & 21,158 & 14,112 & 10,589 & 8,475  \\
\bottomrule
\end{tabular}}
\caption{The amount of storage (bits) for different composite sample sizes on various datasets.}
\label{tab:table5}
\end{table}

\section{Conclusion}
In this study, we introduce a novel reversible adversarial attack framework for audio privacy protection. IO-RAE leverages LLMs to generate target attack phrases and employs cumulative signal techniques to seamlessly integrate noise. IO-RAE demonstrates excellent attack and recovery performance and has been tested against Google's commercial black-box model, achieving a 100\% success rate in non-target attacks. In the future, we plan to explore high-transferability target attack methods in black-box scenarios to enhance its misleading capabilities, thereby achieving more practical and effective audio data protection.

\section{Acknowledgments}
This work was supported in part by the Xiamen Research Project for the Natural Science Foundation of Xiamen,China
(3502Z202472028), the Xiamen Science and Technology Plan Project (3502Z20231042), the Fundamental Research Funds for the Central Universities(1082204112364) and the Science and Technology Development Fund, Macau SAR, under Grant 0193/2023/RIA3 and 0079/2025/AFJ.
\bibliography{Bibliography-File}
\end{document}